\documentclass[8.5 pt,twoside]{article}

\usepackage[super,sort&compress,comma]{natbib} 
\usepackage{times,mathptm}
\usepackage{sectsty}
\usepackage{balance} 
\usepackage[text={11.3cm,20.4cm},centering]{geometry} 

\usepackage{graphicx} 
\usepackage{lastpage}
\usepackage[format=plain,justification=raggedright,singlelinecheck=false,font=small,labelfont=bf,labelsep=space]{caption} 
\usepackage{fancyhdr}
\begin{document}

 \newcommand{\apj}{ApJ} 
 \newcommand{\apss}{Ap\&SS} 
 \newcommand{\apjl}{ApJL} 
 \newcommand{\aap}{A\&A} 
 \newcommand{\araa}{Annu. Rev. Astron. Astrophys.} 
 \newcommand{\icarus}{Icarus} 
 \newcommand{\gca}{Geochim. Cosmochim. Acta} 
 \newcommand{\nat}{Nature} 
 \newcommand{\apjs}{ApJ Supplement Series}

\thispagestyle{plain}
\fancypagestyle{plain}{
\fancyhead[L]{\includegraphics[height=8pt]{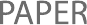}}
\fancyhead[C]{\hspace{-1cm}\includegraphics[height=15pt]{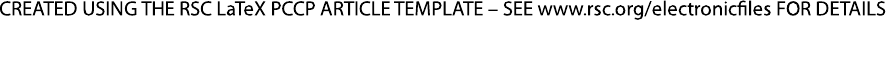}}
\fancyhead[R]{\includegraphics[height=10pt]{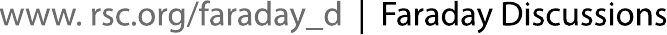}\vspace{-0.2cm}}
\renewcommand{\headrulewidth}{1pt}}
\renewcommand{\thefootnote}{\fnsymbol{footnote}}
\renewcommand\footnoterule{\vspace*{1pt}%
\hrule width 11.3cm height 0.4pt \vspace*{5pt}} 
\setcounter{secnumdepth}{5}

\makeatletter 
\renewcommand{\fnum@figure}{\textbf{Fig.~\thefigure~~}}
\def\subsubsection{\@startsection{subsubsection}{3}{10pt}{-1.25ex plus -1ex minus -.1ex}{0ex plus 0ex}{\normalsize\bf}} 
\def\paragraph{\@startsection{paragraph}{4}{10pt}{-1.25ex plus -1ex minus -.1ex}{0ex plus 0ex}{\normalsize\textit}} 
\renewcommand\@biblabel[1]{#1}            
\renewcommand\@makefntext[1]%
{\noindent\makebox[0pt][r]{\@thefnmark\,}#1}
\makeatother 
\sectionfont{\large}
\subsectionfont{\normalsize} 

\fancyfoot{}
\fancyfoot[LO,RE]{\vspace{-7pt}\includegraphics[height=8pt]{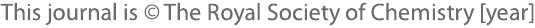}}
\fancyfoot[CO]{\vspace{-7pt}\hspace{5.9cm}\includegraphics[height=7pt]{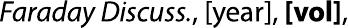}}
\fancyfoot[CE]{\vspace{-6.6pt}\hspace{-7.2cm}\includegraphics[height=7pt]{RF}}
\fancyfoot[RO]{\scriptsize{\sffamily{1--\pageref{LastPage} ~\textbar  \hspace{2pt}\thepage}}}
\fancyfoot[LE]{\scriptsize{\sffamily{\thepage~\textbar\hspace{3.3cm} 1--\pageref{LastPage}}}}
\fancyhead{}
\renewcommand{\headrulewidth}{1pt} 
\renewcommand{\footrulewidth}{1pt}
\setlength{\arrayrulewidth}{1pt}
\setlength{\columnsep}{6.5mm}
\setlength\bibsep{1pt}

\begin{center}
\noindent\LARGE{\textbf{The chemistry of planet-forming regions is not interstellar}}
\end{center}
\vspace{0.6cm}

\noindent\large{\textbf{Klaus M. Pontoppidan$^{\ast}$\textit{$^{a}$} } and \textbf{Sandra M. Blevins\textit{$^{a,b}$}}\vspace{0.5cm}

\noindent\textit{\small{\textbf{Received 15th December 2013, Accepted 4th February 2014}}}

\noindent \textbf{\small{DOI: 10.1039/c3fd00141e}}
\vspace{0.6cm}

\noindent \normalsize{Advances in infrared and submillimeter technology have allowed for detailed observations of the molecular content of the planet-forming regions of protoplanetary disks. In particular, disks around solar-type stars now have growing molecular inventories that can be directly compared with both prestellar chemistry and that inferred for the early solar nebula. The data directly address the old question whether the chemistry of planet-forming matter is similar or different and unique relative to the chemistry of dense clouds and protostellar envelopes. The answer to this question may have profound consequences for the structure and composition of planetary systems. The practical challenge is that observations of emission lines from disks do not easily translate into chemical concentrations. Here, we present a two-dimensional radiative transfer model of RNO 90, a classical protoplanetary disk around a solar-mass star, and retrieve the concentrations of dominant molecular carriers of carbon, oxygen and nitrogen in the terrestrial region around 1\,AU. We compare our results to the chemical inventory of dense clouds and protostellar envelopes, and argue that inner disk chemistry is, as expected, fundamentally different from prestellar chemistry. We find that the clearest discriminant may be the concentration of CO$_2$, which is extremely low in disks, but one of the most abundant constituents of dense clouds and protostellar envelopes. }
\vspace{0.5cm}

\section{Introduction}

\footnotetext{\textit{$^{a}$~Space Telescope Science Institute, 3700
    San Martin Drive, MD 21218, Baltimore, USA.  Tel: +1 410 338 4744; E-mail: pontoppi@stsci.edu}}
\footnotetext{\textit{$^{b}$~Catholic University of America, Physics Department, Washington, DC, 20064, USA}}

Planets are ultimately built from matter that is inherited from diffuse gas in the interstellar medium. The chemical state of this material continuously evolves as it travels a complex path through a dense molecular cloud, protostellar envelope and protoplanetary disk.  The elements in the material are differentially partitioned into gas and solids, and apart from a small contribution from nuclear reactions with stellar and cosmic radiation, their total abundances remain largely unchanged.  Conversely, their molecular carriers, including those carrying the bulk of some elements, may change dramatically along this path, and depending on their volatility the local elemental abundances my be altered by hydrodynamic transport processes\citep{Ciesla06,Oberg11}. This rich history of pre-planetary matter is validated by strong differences in chemical content of primitive chondrites from the 3 AU region of the solar nebula\citep{Scott07}, comets that formed beyond 10s of AU\citep{Mumma11} and protostellar envelopes\citep{Oberg11}.

If planet-forming chemistry, in bulk, is an active, local process, this means that exoplanetary systems, particularly terrestrial planets, are formed from material with chemical properties that are unlike those of their natal envelopes and, likely, cometary systems. This scenario would not be surprising, since, in the solar system, it has long been known that almost all of the matter within a few AU was subjected to temperatures sufficiently high to fundamentally alter its chemistry\citep{Grossman72}. It is therefore of great interest to determine whether strong chemical evolution is occuring in typical protoplanetary disks, not only in the case of refractory material\citep{vanBoekel04} but also for ice-forming molecules such as water, CO$_2$ and volatile organics. A consequence of strong evolution is that prestellar-like chemistry cannot be used directly for input into planet formation models; their input must come from direct observations of protoplanetary disks in cooperation with thermo-chemical modeling applied to physical conditions relevant for inner disks.

In this paper, we consider the recent evidence that the inner planet-forming regions of protoplanetary disks ($<10$\,AU) have a radically different chemistry than that of the dense interstellar medium and the typical protostellar envelope. That the molecular emission from inner disks is indicative of an active chemistry has already been suggested, based on simple zero-dimensional calculations\citep{Carr11}, albeit with significant degeneracies\citep{Salyk11}. As one step toward eliminating such retrieval degeneracies, we are developing more sophisticated models, along with additional constraints. We present preliminary results using two-dimensional radiative transfer models to retrieve molecular abundances from planet-forming regions with greater accuracy than previously possible. Specifically, we will compare the molecular inventory obtained from dense clouds to that of the protoplanetary disk around RNO 90 -- a classical T Tauri star of roughly 1 solar mass.

\subsection{The bulk chemistry of dense clouds in protostellar envelopes}
Prior to the formation of the star and disk, nearly all of the CNO carriers are bound in the solid state, either as ices or in refractory silicates and carbon compounds \citep{Pontoppidan04,Whittet10}. CO, and possibly N$_2$, persists the longest in the gas-phase but even these freeze out at the lowest temperatures. Using infrared absorption spectroscopy toward background or young stars, a very accurate, and nearly complete, estimate of the molecular inventory in dense clouds has been compiled \citep{Boogert08}. Interstellar ices are dominated by H$_2$O and CO but with large contributions CO$_2$ ($\sim 25$\% relative to water), and occasionally CH$_3$OH (up to 25\%, but more typically 5\%). The most abundant ice species are characterized by a remarkably universal chemical signatures, and the concentrations of e.g., CO$_2$, CH$_4$ and various volatile organics do not vary by more than a factor of 2 within nearby quiescent dense molecular clouds \citep{Oberg11}. There are some occasional exceptions to this characteristic ``ice signature'', including CH$_3$OH \citep{Pontoppidan04}, but, generally speaking, dense cloud bulk ice chemistry is quite unique. As we shall see, perhaps CO$_2$, one of the most stable cloud species, offers the strongest evidence that planet-forming chemistry is far from interstellar. 

\subsection{Observations of water and organics in disks with Spitzer}
To discuss similarities and differences in bulk chemistries among distinct stages in protostellar and planetary evolution it is important to have a relatively complete observational inventory to quantify the majority of the elemental constituents.  A significant fraction of missing CNO could be an indication of either uncertain measurements, or that it is sequestered in a chemically and radiatively inactive reservoir. An example of the latter case could be where a gas-phase carrier is observed in a region where a significant part of the element is depleted in ices.  It is well known that gas-phase observations of molecules in cold, dense clouds may provide little information on the bulk chemistry since most species are frozen out as ices. This is particularly a problem if geometric circumstances prevent the use of infrared absorption spectroscopy. For instance, the flattened structure of disks typically do not allow for a line of sight to line up with an infrared background source. However, at 1-2 AU in typical protoplanetary disks, ice condensation is much less of a problem, since gas and dust temperatures are hundreds of K or more, effectively preventing any ice formation.

The Spitzer Space Telescope obtained sensitive mid-infrared spectroscopy of hundreds of protoplanetary disks around young stars across the stellar mass range \citep{Carr08,Salyk08,Pascucci09}. Many of the higher-quality spectra revealed a multitude of lines from many bulk carriers of the volatile elements carbon, oxygen and nitrogen \citep{Pontoppidan10,Carr11}. The species that have so far been seen include H$_2$O, CO, HCN, C$_2$H$_2$, CO$_2$ and OH. The lines come from rovibrational states, or high rotational states in the case of water, with upper level energies ranging from a few hundred K to a few thousand K \citep{Meijerink09}. Furthermore, the wavelength coverage was broad and provided strict upper limits on most molecules potentially carrying a significant fraction of the total elemental abundance of C, N and O. With these data sets it became possible to begin constructing a complete chemical inventory of planet-forming regions \citep{Bast13}. 

\subsection{Previously derived inner disk concentrations}
One challenge is that the Spitzer spectra do not resolve the individual lines, leading to strong line blending. In the analysis of the Spitzer spectra, sophisticated models were often considered over-powered, and relative molecular concentrations were therefore calculated using highly simplified ``slab models'', in which the disk is modeled as a dust-less slab of gas at rest with a single kinetic temperature and level populations in thermodynamic equilibrium (TE). While these assumptions unlikely to hold for protoplanetary disks, with the possible exception of TE, such models are usually able to reproduce the observed spectra of organics and water, at least over a limited wavelength range. Indeed, most slab models are in fact degenerate resulting in a wide range of allowed concentrations. 

Carr \& Najita 2008\cite{Carr08} used a slab model in which they allowed all of the parameters, a single temperature (T), column density (N), and surface emitting area (A), to be free. They found CO/H$_2$O$\sim$1 and HCN/H$_2$O of a few percent. Significantly, this required the emitting area of HCN to be much smaller than that of H$_2$O, although the temperatures were about the same ($\sim 600\,$K). CO$_2$ was also found to have a smaller emitting area than that of H$_2$O but at a much lower temperature of $\sim 350\,$K and a wide range of allowed concentrations, up to 15\% but as low as 0.2\%. Salyk et al. 2011\cite{Salyk11} did not allow all parameters to be free but fixed the emitting areas to that of water essentially assuming that the chemistry remains constant throughout the inner few AU of the disks resulting in similar CO/H$_2$O ratios of $\sim 1$, but much lower HCN/H$_2$O ratios of 0.1\% or less. Further, for CO$_2$ the concentration was as low as 0.1\%, but also found to be highly degenerate.

While arriving at different results the two studies are actually in general and quantitative agreement in that smaller emitting areas lead to higher column densities and vice versa. One problem is that in both cases there is no guarantee that the derived combination of temperatures and emitting areas are consistent with a disk temperature structure. This is the aspect that we now attempt to address; if a realistic disk structure is imposed will this eliminate some of the degenerate parameter space, and allow us to measure inner disk chemistries with precision high enough to determine whether they are different from those measured in protostellar envelopes?

\section{Concentration retrieval using two-dimensional radiative transfer models}

One possible way to limit the degeneracies of slab models is to require that all observed lines from all species are reproduced by a single two-dimensional disk structure in Keplerian rotation, with a self-consistent temperature structure. To this end, we use a two-dimensional radiative transfer model to retrieve the molecular concentrations from infrared spectroscopic observations to higher levels of accuracy than previously achieved with single-slab models. In particular, the higher-order model yields estimates for the absolute local concentration in the disk photospheres ($n(X)/n(H)$). The modeling procedure is based on the commonly used continuum radiative transfer code RADMC \citep{Dullemond04}, along with the line raytracer, RADLite \citep{Pontoppidan09}. 

There are still a number of simplifying assumptions in order to keep the problem contained. For instance, we set the gas temperature to that of the dust which is a good approximation for the hydrogen column density of the molecular layer at mid-infrared wavelengths ($>10^{22}$)\citep{Najita11}. For atomic species and highly optically thick transitions probing higher layers, this approximation would likely be less valid.  Similarly, we assume that all molecular level populations, except for CO, are in thermodynamic equilibrium. This is generally justified by the high densities of the gas that form the lines ($n\sim 10^8-10^{12}\,\rm cm^{-3}$), although future studies should include a general non-LTE treatment to verify this assumption. The non-LTE calculations carried out for CO are based on the escape probability methodology of Woitke et al. 2009\citep{Woitke09} and are described in greater detail by Lockwood et al., in prep.  We also assume a constant gas-to-dust ratio that was increased from the canonical value of 100 to 500 to include some dust settling. Since the disk structure is fixed by dust emission (Blevins et al., in prep), a different choice of gas-to-dust ratio will simply scale the total disk mass as $\sim g2d/500$ and the absolute chemical concentrations as $\sim 500/g2d$. Relative concentrations are not affected. 

\subsection{Modeling procedure}

For a given disk, we begin by fitting the two-dimensional dust distribution to the observed spectral energy distribution (SED), after correcting for any interstellar extinction. The free parameters include the disk mass, the shape of the disk surface (the flaring index), and the height and radius of the disk. Optimization of these parameters results in a dust temperature and mean intensity spectrum at every location in the disk. Given the dust temperature model and a molecular concentration at every $(R,z)$ point in the disk, line spectra are rendered with the raytracer, RADLite. In other investigations, the local molecular concentrations are often generated using a chemical model \citep{Woitke09} but since we are interested in determining what the concentrations are, independent of any model, we parameterize the spatial concentration distribution using an inner and outer abundance, along with a transition radius. The transition radius could be synonymous with a snow line -- the radius at which water or another molecule freezes out -- or it could be a pure chemical boundary.  The same methods were used to model the water vapor emission for TW Hya \citep{Zhang13}.

Since we only fit two parameters for each molecule (the outer abundance can typically be set to 0 in this context), we can find the best solution by generating a grid of models on those two parameters and minimizing the difference between the model spectra and the data after matching their continua using an additive straight line to correct for any minor differences ($\sim 10$\% in this case). Note that RADLite does carry out a self-consistent calculation of the dust continuum, but matching the model continuum to that of the broad band data to much better than 10\%  requires a great deal of fine tuning with little effect on the conclusions. Nevertheless, the immediate advantage of using a two-dimensional disk model, constrained by continuum observations, is that the temperature of the gas forming molecular lines is no longer a free parameter, as it is when using an otherwise unconstrained slab model. Further, the spectra due to each separate molecule are linked because they ultimately have to be produced by the same disk model. Using a slab model there are no such links, and each molecule is completely independent of another. 

Altogether, we simultaneously model H$_2$O, CO, HCN, C$_2$H$_2$ and CO$_2$, accounting for about 2000 individually rendered lines and line-images in the covered spectral ranges. 
 
\subsection{Example disk: RNO 90}

Protoplanetary disks with strong molecular emission are common and since slab models produce very similar results for different sources \citep{Salyk11}, we can choose any one as an example. In the future, more elaborate studies may include larger disk samples. For this paper we chose RNO 90, a young solar-mass star (G5) with strong water vapor emission in the mid-infrared.  RNO 90 is located near (but not inside) the LDN 43 dark cloud roughly 10 degrees from the well-known $\rho$ Ophiuchus young stellar cluster at a distance generally assumed to be that of the cluster of 125\,pc \citep{Loinard08,Mamajek08}.  The disk is optically thick in the infrared at all radii, and there have been no detections of structural ``complications'' as seen in more evolved disks, such as TW Hya \citep{Akeson11}, although RNO 90 is also not as thoroughly observed. The geometric parameters of the RNO 90 disk are summarized in Table \ref{tbl:RNO90_parameters}. 
 
\begin{table}[ht]
\small
\centering
  \caption{~Two-dimensional model parameters for RNO 90}
  \label{tbl:RNO90_parameters}
  \begin{tabular}{lll} 
    \hline
    Structural parameters && \\
    \hline
    Stellar mass & $M_*$ & $1.5\,M_{\odot}$\\
    Disk mass& $M_{\rm disk}$ & $1.5\times 10^{-2}\,M_{\odot}$\\
    Gas-to-dust ratio& $M_{\rm gas}$/$M_{\rm dust}$& 500\\
    Outer radius& $R_{\rm out}$ & 150\,AU\\
    Flaring index& $\alpha$ & 0.06\\
    Dust sublimation temperature & $T_{\rm sub}$ & 1800\,K \\
    \hline
    H$_2$O inner concentration &n(H$_2$O)/n(H)&$2.5\times 10^{-3}$\\
    H$_2$O transition radius &R(H$_2$O)&4\,AU\\
    H$_2$O outer concentration &n(H$_2$O)/n(H)& 0 \\
    CO inner concentration &n(CO)/n(H)&$7.5\times 10^{-5}$\\
    CO transition radius &R(CO)&4\,AU\\
    CO outer concentration &n(CO)/n(H)& 0\\
    C$_2$H$_2$ inner concentration &n(C$_2$H$_2$)/n(H)&$10^{-5}$\\
    C$_2$H$_2$ transition radius &R(C$_2$H$_2$)& 0.2\,AU\\
    C$_2$H$_2$ outer concentration &n(C$_2$H$_2$)/n(H)&0 \\
    HCN inner concentration &n(HCN)/n(H)&$7.5\times 10^{-6}$\\
    HCN transition radius &R(HCN)& 1.3\,AU\\
    HCN outer concentration &n(HCN)/n(H)&$6.5\times 10^{-7}$\\
    CO$_2$ inner concentration &n(CO$_2$)/n(H)&$2.5\times 10^{-7}$\\
    CO$_2$ transition radius &R(CO$_2$)& 4\,AU\\
    CO$_2$ outer concentration &n(CO$_2$)/n(H)&0\\
    \hline
  \end{tabular}
\end{table}

\subsection{Observations}
RNO 90 has one of the highest quality mid-infrared molecular spectra available \citep{Pontoppidan10}, and has an increasing amount of geometric and structural information available, including submillimeter imaging \citep{Andrews07}, Herschel spectroscopy \citep{Podio13}, as well as high resolution CO spectroscopy and spectro-astrometry \citep{Pontoppidan11}.  Specifically, we use the 10-40\,$\mu$m Spitzer spectrum\citep{Pontoppidan10} (resolving power of $500\,\rm km~s^{-1}$) in combination with the high resolution spectrum of the 4.7\,$\mu$m CO rovibrational band obtained with CRIRES (resolving power of $\sim 3\,\rm km~s^{-1}$) on the Very Large Telescope\citep{Pontoppidan11}. The CO spectrum is highly complementary to the Spitzer spectrum, since the former fully resolved the line profiles and velocity structure, while the latter fails to resolve most individual lines. The details of the data reduction can be found in Pontoppidan et al. 2010\citep{Pontoppidan10}. 
\begin{table}[ht]
\footnotesize
\centering
  \caption{~Comparison of observed relative molecular concentrations using different retrieval methods}
  \label{tbl:RNO90_concentrations}
  \begin{tabular}{lllll} 
    \hline
    & & Carr \& Najita 2008, 2011 & Salyk et al. 2011 & RADLite\\
     Model type & Band & Slab & Slab with fixed area & 2D dust+gas \\
    \hline
    n(CO)/n(H$_2$O) & $\nu_1$ & $0.63\pm 0.2$  & $0.79$ & $0.03$ \\
    n(C$_2$H$_2$)/n(H$_2$O) &$\nu_5$  & $(1.0\pm 0.5)\times 10^{-2}$ & $2.5\times 10^{-4}$ &$4\times 10^{-3}$ \\
    n(HCN)/n(H$_2$O) & $\nu_2$ & $(8\pm 5) \times 10^{-2}$ & $1.6\times 10^{-3}$ & $3\times 10^{-3}$\\
    n(CO$_2$)/n(H$_2$O) &$\nu_2$ & $(8\pm 7\times 10^{-2}$ & $8\times 10^{-4}$  & $1\times 10^{-4}$\\
    \hline
  \end{tabular}
\end{table}

\section{The observed inner disk chemistry of RNO 90}

The infrared spectrum of RNO 90 is typical in that it shows strong emission lines from, at least, CO, H$_2$O, OH, C$_2$H$_2$, HCN and CO$_2$. We first fit a water model to the many water lines in the 10-16\,$\mu$m region. There are sufficient numbers of water lines to fit a more detailed radial structure, but to limit the scope of this paper, we use a simple step function with a sharp phase transition at the snow line. For RNO 90, this occurs at 2\,AU in the mid-plane and roughly at 4\,AU nearer to the surface. Note that most of the lines considered have high enough upper level energies that they predominantly trace the region well inside 4\,AU. Thus, the water model is not sensitive to the exact choice of boundary. Once the water model parameters are fixed, we fit the other molecular species, one at a time, to their major spectral bands as indicated in Figures \ref{fgr:RNO90_spitzermodel} and \ref{fgr:RNO90_criresmodel}. The absolute concentrations are listed in Table \ref{tbl:RNO90_parameters}, and the relative (to water) concentrations are shown in Table \ref{tbl:RNO90_concentrations}, where they are compared to the column density values from different slab models. The total integrated molecular column densities as functions of disk radius are shown in Figure \ref{fgr:column_densities}.

\begin{figure}[ht]
\centering
\includegraphics[width=12cm]{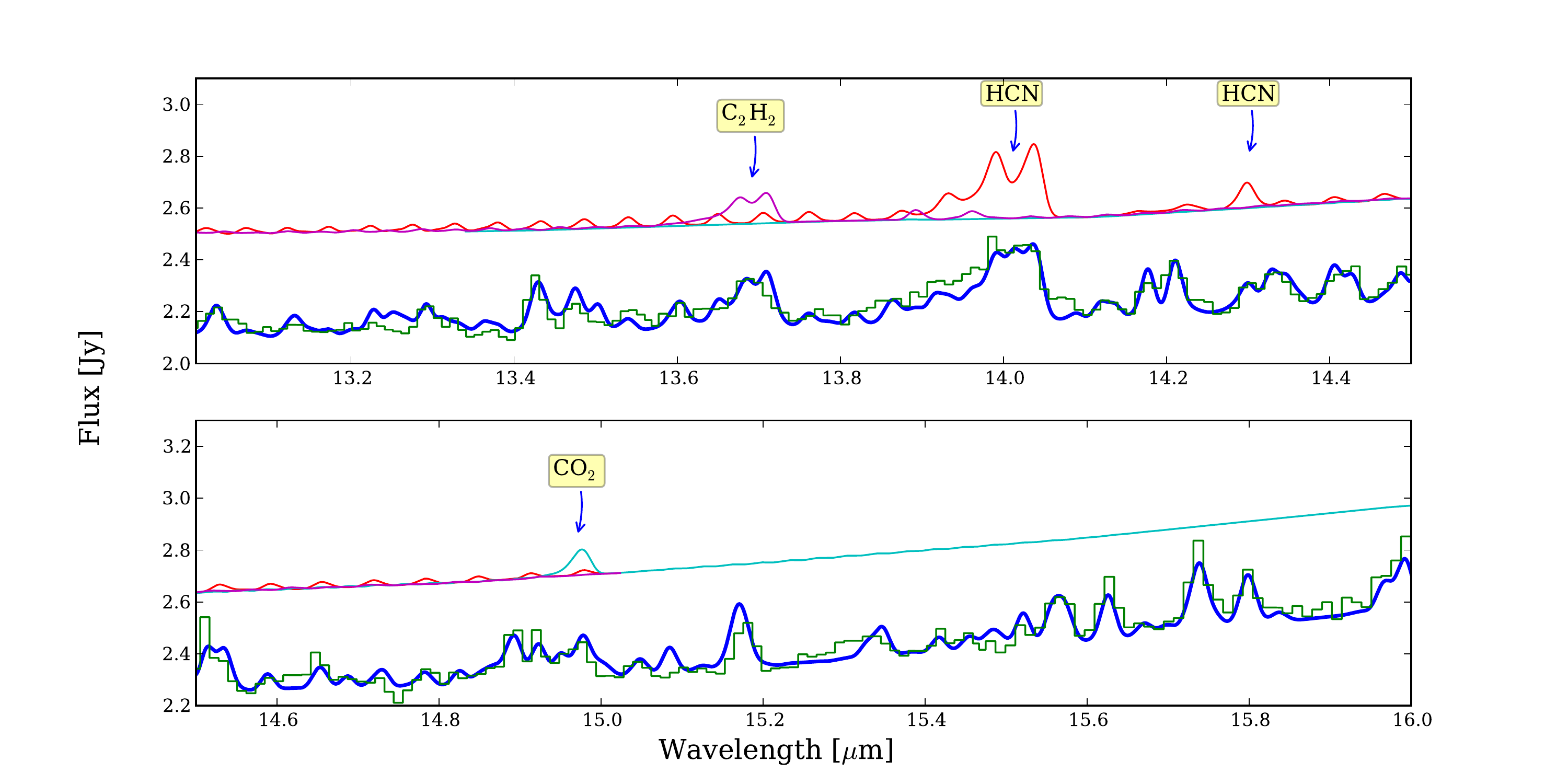}
\caption{Best-fitting two-dimensional model of the Spitzer mid-infrared spectrum of the RNO 90 protoplanetary disk. The continuum of the model has been additively adjusted by $\sim 10$\% to match the observations exactly.The total spectrum is overplotted on the data, while model spectra of the individual species (HCN, C$_2$H$_2$ and CO$_2$) are offset, for clarity. Unmarked model features are due to water. }
\label{fgr:RNO90_spitzermodel}
\end{figure}

\begin{figure}[ht]
\centering
\includegraphics[width=12cm]{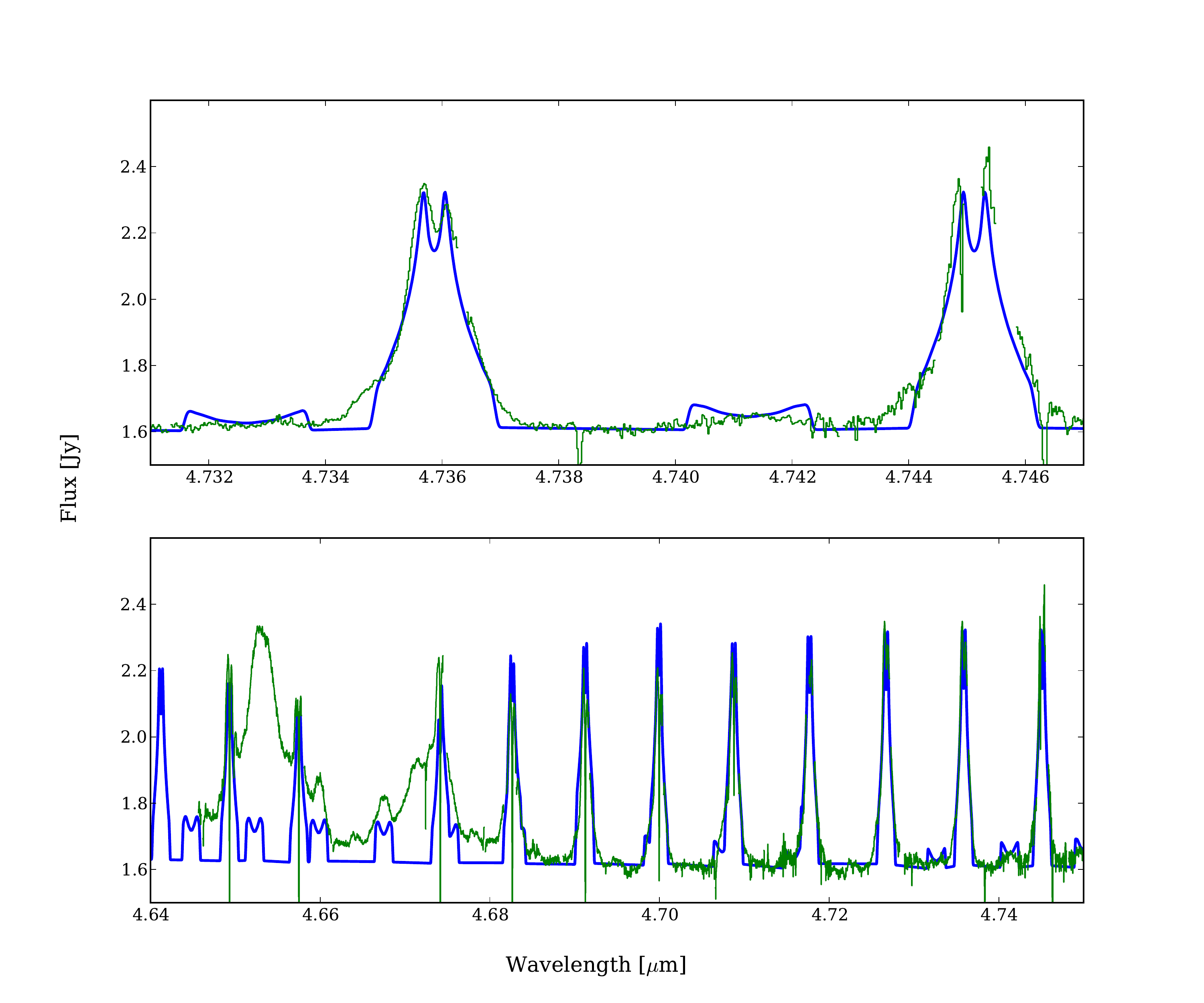}
\caption{As in Figure \ref{fgr:RNO90_spitzermodel}, but for the rovibrational CO fundamental band, as observed with CRIRES. The double-peaked line structure is due to the Keplerian rotation of the disk. The top panel shows the line profiles, while the lower panel shows a wider view of the CO fundamental spectrum. The broad lines at 4.655 and 4.673\,$\mu$m are due to atomic hydrogen.}
\label{fgr:RNO90_criresmodel}
\end{figure}

\subsection{CO}
The CO rovibrational lines, which are fully velocity-resolved by CRIRES, can be fit with little to no degeneracy. Their profiles and strengths are matched well by the structural model of RNO 90, which supports our choice of inner disk structure (see Figure \ref{fgr:RNO90_criresmodel}). One departure is that the $v=2-1$ line profiles are more double-peaked than the data.  This may likely be corrected by adopting a smooth transition of the inner disk edge, rather than the sharp inner edge used by RADMC. The inner disk CO/H model concentration is $7.5\times 10^{-5}$, which is very close to the canonical value supported by thermo-chemical models \citep{Bertelsen14}. This also matches the abundance of dense clouds in the interstellar medium \citep{Dickman78}. That the CO concentrations are so close to canonical supports the absolute values derived for other chemical species formed in the same gas. 

\subsection{HCN}
In the case of HCN, which has a spectrally resolved Q branch band structure, the model fit is definite. Conversely, the slab models are degenerate between optically thick, high column density parameters and optically thin, high temperature parameters\cite{Salyk11}. Since the temperature distribution is fixed in our two-dimensional model, we find that the HCN emission must be optically thick. To reproduce the total band strength we are driven toward a model in which the HCN emission is much closer to the star than the water emission. Further, the HCN bands (near 14.0 and 14.28\,$\mu$m) cannot be fit simultaneously with a constant abundance model. Thus, in the case of HCN, we impose a boundary around 1.3\,AU, with a high concentration on the inside of the boundary, and a slightly lower concentration outside; both regions contribute significantly to the combined HCN emission spectrum. 

\subsection{C$_2$H$_2$}
C$_2$H$_2$ behaves like HCN in RNO 90, but seems more extreme. The band shape seems to requires an extremely high concentration in the innermost region of the disk (the current model uses 0.2\,AU). The band is not as well resolved or as strong as the HCN band in RNO 90, so the C$_2$H$_2$ parameters are still uncertain, even with the two-dimensional model. 

\subsection{CO$_2$}
CO$_2$ is a particularly interesting case. The CO$_2$ Q branch is not spectrally resolved, but it can be fit sufficiently with a constant concentration defined throughout the disk (out to the water transition radius of 4\,AU). The best fitting concentration has a very low value of $10^{-4}$ relative to water. Analogous to the slab models, the maximum concentration can be increased by introducing a radial chemical boundary, beyond which the concentration drops to 0 (or a very low value). However, even if the transition radius is set very low, there will still be a large area of the disk, where the n(CO$_2$)/n(H$_2$O) concentration is several orders of magnitude below the dense cloud/protostellar value of 0.25. It is very difficult to explain this if the inner disk volatiles are predominantly supplied by evaporating icy bodies that were originally formed at low temperatures in the outer disk or in a protostellar envelope. 

\begin{figure}[ht]
\centering
\includegraphics[width=12cm]{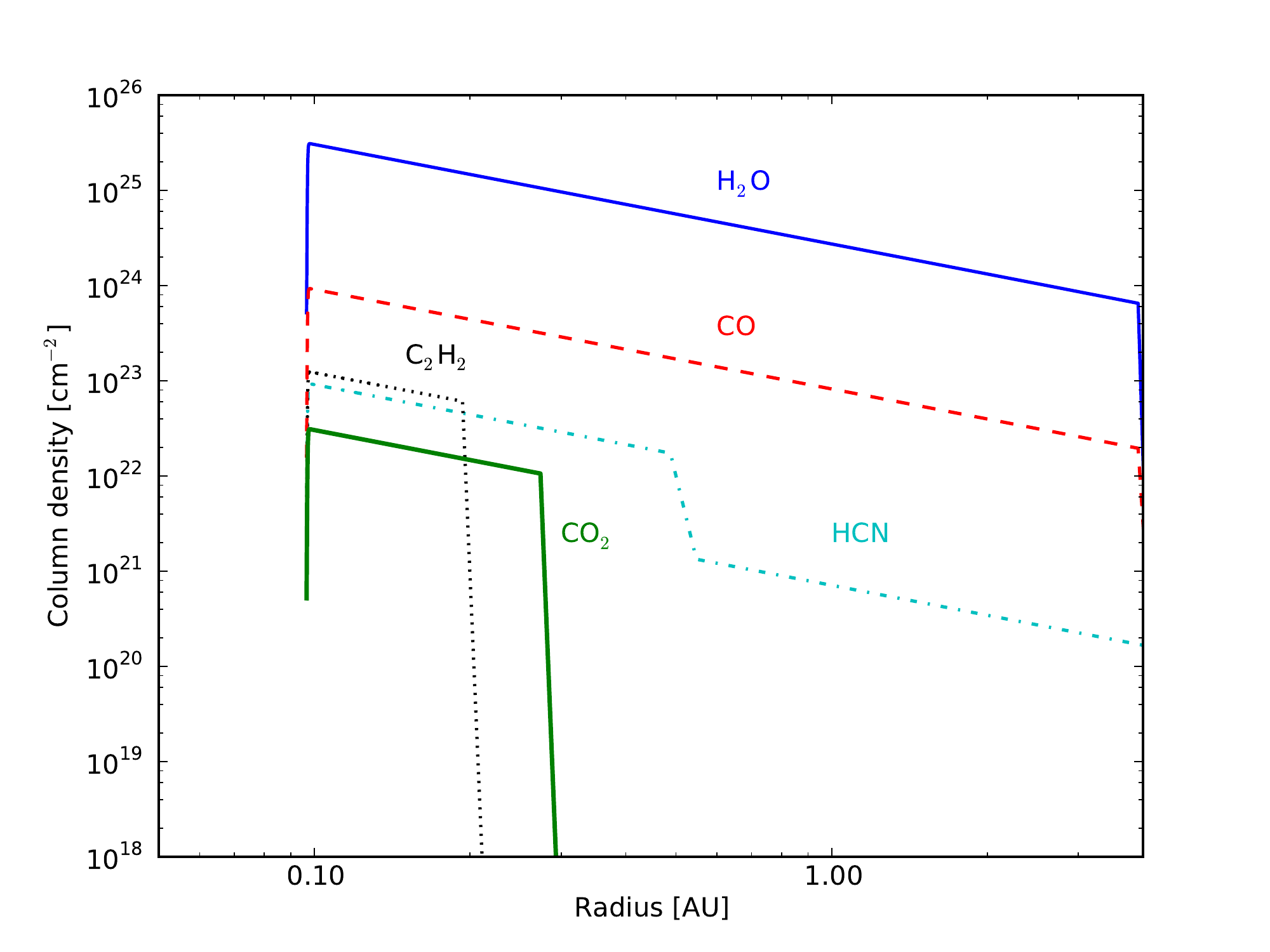}
\caption{Vertically integrated molecular column densities as a function of disk radius. The observations trace only a small fraction ($\sim$0.1\%) of the total column included in the disk, so the values of the figure should be considered an extrapolation.}
\label{fgr:column_densities}
\end{figure}

\section{Discussion}
It is clear from the derived concentrations that they do not resemble those of prestellar ices. In particular CO$_2$ has much lower concentrations in disks. This is in particularly stark contrast to the nearly universal (within a factor 2) concentration of 25\% CO$_2$ relative to water in low-mass star forming regions \citep{Pontoppidan08}. 

Conversely, thermochemical models predict relatively low CO$_2$ concentrations\cite{Willacy09,Najita11} in the planet-forming regions of protoplanetary disks. Indeed, the high concentration of CO$_2$ (and other species) in interstellar ices has long been seen as strong evidence for the action of surface chemistry\citep{dHendecourt85}. The model concentrations of HCN also tend to be low, with most of the nitrogen carried by N$_2$ in the inner disk.  As can be seen from the absolute concentrations (relative to H) listed in Table \ref{tbl:RNO90_absolute}, some of the observed values are consistent with gas-phase thermochemical models. The exceptions are water, which is more abundant by as much as an order of magnitude, and C$_2$H$_2$, which is observed to be several orders of magnitude more abundant. 

One potential way of explaining the increase of the water concentration (beyond what is possible with the solar abundance of oxygen -- $\sim 4\times 10^{-4}$ from \cite{Grevesse10}) is if the water has been been preferentially enriched by advection (inwards migration of icy bodies against the disk pressure gradient) \citep{Ciesla06}. Such bodies might be preferentially water-rich, if they formed outside the water snow line, but inside the CO$_2$, HCN and CO lines. This was proposed by Banzatti et al. 2013\citep{Banzatti13} to explain the high water column densities suggested by some slab models. Alternately, we may have underestimated the gas-to-dust ratio. Increasing this will place higher column densities into the line formation region, and therefore decrease the absolute concentrations, bringing water into the canonical range, but suppressing CO$_2$ and HCN below the thermochemical models. 

It is arguably less surprising that the relative concentrations indicate a very different chemistry from those of ices, even if material ultimately originate in evaporating icy bodies. This could be related to the extremely short chemical time scales at the relevant temperatures and densities. \cite{Semenov11} note that the timescale for gas-phase chemistry in the molecular layer of the inner disk is extremely fast ($<1\,$ hour inside 10\,AU). The same region is somewhat shielded from direct irradiation by ultra-violet photons from the central star and accretion shock that photochemistry timescales are likely somewhat longer ($>1\,$year). 

Another aspect that we have not addressed by modeling a single source is the known strong dependence of the chemistry on stellar type. Hot young stars (Herbig Ae) have disks with weak or absent molecular infrared emission\citep{Pontoppidan10}, while disks around very low-mass stars have weak water, but strong organic lines\citep{Pascucci13}. A common two-dimensional model, such as that presented here, is needed to derive comparable chemical concentrations, and get a complete picture of protoplanetary chemistry across the stellar mass range. 

It seems that whichever way we look at the data, the chemistry of inner disks does not appear to be reflective of a protostellar origin. We would not expect to see strong water emission at the lower canonical water abundance, but would have expected to see a very strong CO$_2$ feature. Other species would also show strong features at prestellar abundances, including CH$_3$OH and NH$_3$, but are yet to be seen\citep{Mandell08}. However, it is very likely that new and interesting chemistry will appear in the mid-infrared once we have space-based spectroscopy at higher spectral resolution (such as will be available with the James Webb Space Telescope). With such data we will be in a better position to understand what indeed drives planet-forming chemistry, knowing that inner disks are highly active chemical factories. 

\begin{table}[ht]
\footnotesize
\centering
  \caption{~Comparison of the observed absolute molecular concentrations with thermochemical disk models at 1\,AU}
  \label{tbl:RNO90_absolute}
  \begin{tabular}{llll} 
    \hline
     & Willacy \& Woods 2009\citep{Willacy09} & Najita et al. 2011\citep{Najita11} & RADLite (this paper)\\
    \hline
    n(H$_2$O) &$1.4\times 10^{-4}$& $10^{-6}-10^{-4}$ & $5\times 10^{-3}$\\
    n(CO) &$2.2\times 10^{-5}$& $10^{-4}$& $7.5\times 10^{-5}$\\
    n(C$_2$H$_2$)&$7.2\times 10^{-7}$ & $10^{-10}-10^{-8}$ & $1\times 10^{-5}$$^a$\\
    n(HCN)  &$5.1\times 10^{-6}$& $10^{-7}$ & $5\times 10^{-6}$ \\
    n(CO$_2$)  &$1.4\times 10^{-6}$& $10^{-10}-10^{-7}$ & $2.5\times 10^{-7}$\\
    \hline
  \end{tabular}
  \begin{itemize}
    \item[$^a$] The C$_2$H$_2$ concentration is valid inside 0.2\,AU. Outside of this, it is at least an order of magnitude lower.
    \end{itemize}
\end{table}

\section*{Acknowledgements}
This work is based in part on observations made with the Spitzer Space Telescope, which is operated by the Jet Propulsion Laboratory, California Institute of Technology under a contract with NASA. Support for this work was provided by NASA through an award issued by JPL/Caltech. Based in part on observations made with ESO Telescopes at the La Silla Paranal Observatory under programme ID 179.C-0151. 

\footnotesize{
\bibliography{faraday}

\providecommand*{\mcitethebibliography}{\thebibliography}
\csname @ifundefined\endcsname{endmcitethebibliography}
{\let\endmcitethebibliography\endthebibliography}{}
\begin{mcitethebibliography}{38}
\providecommand*{\natexlab}[1]{#1}
\providecommand*{\mciteSetBstSublistMode}[1]{}
\providecommand*{\mciteSetBstMaxWidthForm}[2]{}
\providecommand*{\mciteBstWouldAddEndPuncttrue}
  {\def\EndOfBibitem{\unskip.}}
\providecommand*{\mciteBstWouldAddEndPunctfalse}
  {\let\EndOfBibitem\relax}
\providecommand*{\mciteSetBstMidEndSepPunct}[3]{}
\providecommand*{\mciteSetBstSublistLabelBeginEnd}[3]{}
\providecommand*{\EndOfBibitem}{}
\mciteSetBstSublistMode{f}
\mciteSetBstMaxWidthForm{subitem}
{(\emph{\alph{mcitesubitemcount}})}
\mciteSetBstSublistLabelBeginEnd{\mcitemaxwidthsubitemform\space}
{\relax}{\relax}

\bibitem[{Ciesla} and {Cuzzi}(2006)]{Ciesla06}
F.~J. {Ciesla} and J.~N. {Cuzzi}, \emph{\icarus}, 2006, \textbf{181},
  178--204\relax
\mciteBstWouldAddEndPuncttrue
\mciteSetBstMidEndSepPunct{\mcitedefaultmidpunct}
{\mcitedefaultendpunct}{\mcitedefaultseppunct}\relax
\EndOfBibitem
\bibitem[{{\"O}berg} \emph{et~al.}(2011){{\"O}berg}, {Boogert}, {Pontoppidan},
  {van den Broek}, {van Dishoeck}, {Bottinelli}, {Blake}, and {Evans}]{Oberg11}
K.~I. {{\"O}berg}, A.~C.~A. {Boogert}, K.~M. {Pontoppidan}, S.~{van den Broek},
  E.~F. {van Dishoeck}, S.~{Bottinelli}, G.~A. {Blake} and N.~J. {Evans}, II,
  \emph{\apj}, 2011, \textbf{740}, 109\relax
\mciteBstWouldAddEndPuncttrue
\mciteSetBstMidEndSepPunct{\mcitedefaultmidpunct}
{\mcitedefaultendpunct}{\mcitedefaultseppunct}\relax
\EndOfBibitem
\bibitem[{Scott}(2007)]{Scott07}
E.~R.~D. {Scott}, \emph{Annual Review of Earth and Planetary Sciences}, 2007,
  \textbf{35}, 577--620\relax
\mciteBstWouldAddEndPuncttrue
\mciteSetBstMidEndSepPunct{\mcitedefaultmidpunct}
{\mcitedefaultendpunct}{\mcitedefaultseppunct}\relax
\EndOfBibitem
\bibitem[{Mumma} and {Charnley}(2011)]{Mumma11}
M.~J. {Mumma} and S.~B. {Charnley}, \emph{\araa}, 2011, \textbf{49},
  471--524\relax
\mciteBstWouldAddEndPuncttrue
\mciteSetBstMidEndSepPunct{\mcitedefaultmidpunct}
{\mcitedefaultendpunct}{\mcitedefaultseppunct}\relax
\EndOfBibitem
\bibitem[{Grossman}(1972)]{Grossman72}
L.~{Grossman}, \emph{\gca}, 1972, \textbf{36}, 597--619\relax
\mciteBstWouldAddEndPuncttrue
\mciteSetBstMidEndSepPunct{\mcitedefaultmidpunct}
{\mcitedefaultendpunct}{\mcitedefaultseppunct}\relax
\EndOfBibitem
\bibitem[{van Boekel} \emph{et~al.}(2004){van Boekel}, {Min}, {Leinert},
  {Waters}, {Richichi}, {Chesneau}, {Dominik}, {Jaffe}, {Dutrey}, {Graser},
  {Henning}, {de Jong}, {K{\"o}hler}, {de Koter}, {Lopez}, {Malbet}, {Morel},
  {Paresce}, {Perrin}, {Preibisch}, {Przygodda}, {Sch{\"o}ller}, and
  {Wittkowski}]{vanBoekel04}
R.~{van Boekel}, M.~{Min}, C.~{Leinert}, L.~B.~F.~M. {Waters}, A.~{Richichi},
  O.~{Chesneau}, C.~{Dominik}, W.~{Jaffe}, A.~{Dutrey}, U.~{Graser},
  T.~{Henning}, J.~{de Jong}, R.~{K{\"o}hler}, A.~{de Koter}, B.~{Lopez},
  F.~{Malbet}, S.~{Morel}, F.~{Paresce}, G.~{Perrin}, T.~{Preibisch},
  F.~{Przygodda}, M.~{Sch{\"o}ller} and M.~{Wittkowski}, \emph{\nat}, 2004,
  \textbf{432}, 479--482\relax
\mciteBstWouldAddEndPuncttrue
\mciteSetBstMidEndSepPunct{\mcitedefaultmidpunct}
{\mcitedefaultendpunct}{\mcitedefaultseppunct}\relax
\EndOfBibitem
\bibitem[{Carr} and {Najita}(2011)]{Carr11}
J.~S. {Carr} and J.~R. {Najita}, \emph{\apj}, 2011, \textbf{733}, 102\relax
\mciteBstWouldAddEndPuncttrue
\mciteSetBstMidEndSepPunct{\mcitedefaultmidpunct}
{\mcitedefaultendpunct}{\mcitedefaultseppunct}\relax
\EndOfBibitem
\bibitem[{Salyk} \emph{et~al.}(2011){Salyk}, {Pontoppidan}, {Blake}, {Najita},
  and {Carr}]{Salyk11}
C.~{Salyk}, K.~M. {Pontoppidan}, G.~A. {Blake}, J.~R. {Najita} and J.~S.
  {Carr}, \emph{\apj}, 2011, \textbf{731}, 130\relax
\mciteBstWouldAddEndPuncttrue
\mciteSetBstMidEndSepPunct{\mcitedefaultmidpunct}
{\mcitedefaultendpunct}{\mcitedefaultseppunct}\relax
\EndOfBibitem
\bibitem[{Pontoppidan} \emph{et~al.}(2004){Pontoppidan}, {van Dishoeck}, and
  {Dartois}]{Pontoppidan04}
K.~M. {Pontoppidan}, E.~F. {van Dishoeck} and E.~{Dartois}, \emph{\aap}, 2004,
  \textbf{426}, 925--940\relax
\mciteBstWouldAddEndPuncttrue
\mciteSetBstMidEndSepPunct{\mcitedefaultmidpunct}
{\mcitedefaultendpunct}{\mcitedefaultseppunct}\relax
\EndOfBibitem
\bibitem[{Whittet}(2010)]{Whittet10}
D.~C.~B. {Whittet}, \emph{\apj}, 2010, \textbf{710}, 1009--1016\relax
\mciteBstWouldAddEndPuncttrue
\mciteSetBstMidEndSepPunct{\mcitedefaultmidpunct}
{\mcitedefaultendpunct}{\mcitedefaultseppunct}\relax
\EndOfBibitem
\bibitem[{Boogert} \emph{et~al.}(2008){Boogert}, {Pontoppidan}, {Knez},
  {Lahuis}, {Kessler-Silacci}, {van Dishoeck}, {Blake}, {Augereau}, {Bisschop},
  {Bottinelli}, {Brooke}, {Brown}, {Crapsi}, {Evans}, {Fraser}, {Geers},
  {Huard}, {J{\o}rgensen}, {{\"O}berg}, {Allen}, {Harvey}, {Koerner}, {Mundy},
  {Padgett}, {Sargent}, and {Stapelfeldt}]{Boogert08}
A.~C.~A. {Boogert}, K.~M. {Pontoppidan}, C.~{Knez}, F.~{Lahuis},
  J.~{Kessler-Silacci}, E.~F. {van Dishoeck}, G.~A. {Blake}, J.-C. {Augereau},
  S.~E. {Bisschop}, S.~{Bottinelli}, T.~Y. {Brooke}, J.~{Brown}, A.~{Crapsi},
  N.~J. {Evans}, II, H.~J. {Fraser}, V.~{Geers}, T.~L. {Huard}, J.~K.
  {J{\o}rgensen}, K.~I. {{\"O}berg}, L.~E. {Allen}, P.~M. {Harvey}, D.~W.
  {Koerner}, L.~G. {Mundy}, D.~L. {Padgett}, A.~I. {Sargent} and K.~R.
  {Stapelfeldt}, \emph{\apj}, 2008, \textbf{678}, 985--1004\relax
\mciteBstWouldAddEndPuncttrue
\mciteSetBstMidEndSepPunct{\mcitedefaultmidpunct}
{\mcitedefaultendpunct}{\mcitedefaultseppunct}\relax
\EndOfBibitem
\bibitem[{Carr} and {Najita}(2008)]{Carr08}
J.~S. {Carr} and J.~R. {Najita}, \emph{Science}, 2008, \textbf{319},
  1504--\relax
\mciteBstWouldAddEndPuncttrue
\mciteSetBstMidEndSepPunct{\mcitedefaultmidpunct}
{\mcitedefaultendpunct}{\mcitedefaultseppunct}\relax
\EndOfBibitem
\bibitem[{Salyk} \emph{et~al.}(2008){Salyk}, {Pontoppidan}, {Blake}, {Lahuis},
  {van Dishoeck}, and {Evans}]{Salyk08}
C.~{Salyk}, K.~M. {Pontoppidan}, G.~A. {Blake}, F.~{Lahuis}, E.~F. {van
  Dishoeck} and N.~J. {Evans}, II, \emph{\apjl}, 2008, \textbf{676},
  L49--L52\relax
\mciteBstWouldAddEndPuncttrue
\mciteSetBstMidEndSepPunct{\mcitedefaultmidpunct}
{\mcitedefaultendpunct}{\mcitedefaultseppunct}\relax
\EndOfBibitem
\bibitem[{Pascucci} \emph{et~al.}(2009){Pascucci}, {Apai}, {Luhman}, {Henning},
  {Bouwman}, {Meyer}, {Lahuis}, and {Natta}]{Pascucci09}
I.~{Pascucci}, D.~{Apai}, K.~{Luhman}, T.~{Henning}, J.~{Bouwman}, M.~R.
  {Meyer}, F.~{Lahuis} and A.~{Natta}, \emph{\apj}, 2009, \textbf{696},
  143--159\relax
\mciteBstWouldAddEndPuncttrue
\mciteSetBstMidEndSepPunct{\mcitedefaultmidpunct}
{\mcitedefaultendpunct}{\mcitedefaultseppunct}\relax
\EndOfBibitem
\bibitem[{Pontoppidan} \emph{et~al.}(2010){Pontoppidan}, {Salyk}, {Blake},
  {Meijerink}, {Carr}, and {Najita}]{Pontoppidan10}
K.~M. {Pontoppidan}, C.~{Salyk}, G.~A. {Blake}, R.~{Meijerink}, J.~S. {Carr}
  and J.~{Najita}, \emph{\apj}, 2010, \textbf{720}, 887--903\relax
\mciteBstWouldAddEndPuncttrue
\mciteSetBstMidEndSepPunct{\mcitedefaultmidpunct}
{\mcitedefaultendpunct}{\mcitedefaultseppunct}\relax
\EndOfBibitem
\bibitem[{Meijerink} \emph{et~al.}(2009){Meijerink}, {Pontoppidan}, {Blake},
  {Poelman}, and {Dullemond}]{Meijerink09}
R.~{Meijerink}, K.~M. {Pontoppidan}, G.~A. {Blake}, D.~R. {Poelman} and C.~P.
  {Dullemond}, \emph{\apj}, 2009, \textbf{704}, 1471--1481\relax
\mciteBstWouldAddEndPuncttrue
\mciteSetBstMidEndSepPunct{\mcitedefaultmidpunct}
{\mcitedefaultendpunct}{\mcitedefaultseppunct}\relax
\EndOfBibitem
\bibitem[{Bast} \emph{et~al.}(2013){Bast}, {Lahuis}, {van Dishoeck}, and
  {Tielens}]{Bast13}
J.~E. {Bast}, F.~{Lahuis}, E.~F. {van Dishoeck} and A.~G.~G.~M. {Tielens},
  \emph{\aap}, 2013, \textbf{551}, A118\relax
\mciteBstWouldAddEndPuncttrue
\mciteSetBstMidEndSepPunct{\mcitedefaultmidpunct}
{\mcitedefaultendpunct}{\mcitedefaultseppunct}\relax
\EndOfBibitem
\bibitem[{Dullemond} and {Dominik}(2004)]{Dullemond04}
C.~P. {Dullemond} and C.~{Dominik}, \emph{\aap}, 2004, \textbf{417},
  159--168\relax
\mciteBstWouldAddEndPuncttrue
\mciteSetBstMidEndSepPunct{\mcitedefaultmidpunct}
{\mcitedefaultendpunct}{\mcitedefaultseppunct}\relax
\EndOfBibitem
\bibitem[{Pontoppidan} \emph{et~al.}(2009){Pontoppidan}, {Meijerink},
  {Dullemond}, and {Blake}]{Pontoppidan09}
K.~M. {Pontoppidan}, R.~{Meijerink}, C.~P. {Dullemond} and G.~A. {Blake},
  \emph{\apj}, 2009, \textbf{704}, 1482--1494\relax
\mciteBstWouldAddEndPuncttrue
\mciteSetBstMidEndSepPunct{\mcitedefaultmidpunct}
{\mcitedefaultendpunct}{\mcitedefaultseppunct}\relax
\EndOfBibitem
\bibitem[{Najita} \emph{et~al.}(2011){Najita}, {{\'A}d{\'a}mkovics}, and
  {Glassgold}]{Najita11}
J.~R. {Najita}, M.~{{\'A}d{\'a}mkovics} and A.~E. {Glassgold}, \emph{\apj},
  2011, \textbf{743}, 147\relax
\mciteBstWouldAddEndPuncttrue
\mciteSetBstMidEndSepPunct{\mcitedefaultmidpunct}
{\mcitedefaultendpunct}{\mcitedefaultseppunct}\relax
\EndOfBibitem
\bibitem[{Woitke} \emph{et~al.}(2009){Woitke}, {Kamp}, and {Thi}]{Woitke09}
P.~{Woitke}, I.~{Kamp} and W.-F. {Thi}, \emph{\aap}, 2009, \textbf{501},
  383--406\relax
\mciteBstWouldAddEndPuncttrue
\mciteSetBstMidEndSepPunct{\mcitedefaultmidpunct}
{\mcitedefaultendpunct}{\mcitedefaultseppunct}\relax
\EndOfBibitem
\bibitem[{Zhang} \emph{et~al.}(2013){Zhang}, {Pontoppidan}, {Salyk}, and
  {Blake}]{Zhang13}
K.~{Zhang}, K.~M. {Pontoppidan}, C.~{Salyk} and G.~A. {Blake}, \emph{\apj},
  2013, \textbf{766}, 82\relax
\mciteBstWouldAddEndPuncttrue
\mciteSetBstMidEndSepPunct{\mcitedefaultmidpunct}
{\mcitedefaultendpunct}{\mcitedefaultseppunct}\relax
\EndOfBibitem
\bibitem[{Loinard} \emph{et~al.}(2008){Loinard}, {Torres}, {Mioduszewski}, and
  {Rodr{\'{\i}}guez}]{Loinard08}
L.~{Loinard}, R.~M. {Torres}, A.~J. {Mioduszewski} and L.~F.
  {Rodr{\'{\i}}guez}, \emph{\apjl}, 2008, \textbf{675}, L29--L32\relax
\mciteBstWouldAddEndPuncttrue
\mciteSetBstMidEndSepPunct{\mcitedefaultmidpunct}
{\mcitedefaultendpunct}{\mcitedefaultseppunct}\relax
\EndOfBibitem
\bibitem[{Mamajek}(2008)]{Mamajek08}
E.~E. {Mamajek}, \emph{Astronomische Nachrichten}, 2008, \textbf{329}, 10\relax
\mciteBstWouldAddEndPuncttrue
\mciteSetBstMidEndSepPunct{\mcitedefaultmidpunct}
{\mcitedefaultendpunct}{\mcitedefaultseppunct}\relax
\EndOfBibitem
\bibitem[{Akeson} \emph{et~al.}(2011){Akeson}, {Millan-Gabet}, {Ciardi},
  {Boden}, {Sargent}, {Monnier}, {McAlister}, {ten Brummelaar}, {Sturmann},
  {Sturmann}, and {Turner}]{Akeson11}
R.~L. {Akeson}, R.~{Millan-Gabet}, D.~R. {Ciardi}, A.~F. {Boden}, A.~I.
  {Sargent}, J.~D. {Monnier}, H.~{McAlister}, T.~{ten Brummelaar},
  J.~{Sturmann}, L.~{Sturmann} and N.~{Turner}, \emph{\apj}, 2011,
  \textbf{728}, 96\relax
\mciteBstWouldAddEndPuncttrue
\mciteSetBstMidEndSepPunct{\mcitedefaultmidpunct}
{\mcitedefaultendpunct}{\mcitedefaultseppunct}\relax
\EndOfBibitem
\bibitem[{Andrews} and {Williams}(2007)]{Andrews07}
S.~M. {Andrews} and J.~P. {Williams}, \emph{\apj}, 2007, \textbf{671},
  1800--1812\relax
\mciteBstWouldAddEndPuncttrue
\mciteSetBstMidEndSepPunct{\mcitedefaultmidpunct}
{\mcitedefaultendpunct}{\mcitedefaultseppunct}\relax
\EndOfBibitem
\bibitem[{Podio} \emph{et~al.}(2013){Podio}, {Kamp}, {Codella}, {Cabrit},
  {Nisini}, {Dougados}, {Sandell}, {Williams}, {Testi}, {Thi}, {Woitke},
  {Meijerink}, {Spaans}, {Aresu}, {M{\'e}nard}, and {Pinte}]{Podio13}
L.~{Podio}, I.~{Kamp}, C.~{Codella}, S.~{Cabrit}, B.~{Nisini}, C.~{Dougados},
  G.~{Sandell}, J.~P. {Williams}, L.~{Testi}, W.-F. {Thi}, P.~{Woitke},
  R.~{Meijerink}, M.~{Spaans}, G.~{Aresu}, F.~{M{\'e}nard} and C.~{Pinte},
  \emph{\apjl}, 2013, \textbf{766}, L5\relax
\mciteBstWouldAddEndPuncttrue
\mciteSetBstMidEndSepPunct{\mcitedefaultmidpunct}
{\mcitedefaultendpunct}{\mcitedefaultseppunct}\relax
\EndOfBibitem
\bibitem[{Pontoppidan} \emph{et~al.}(2011){Pontoppidan}, {Blake}, and
  {Smette}]{Pontoppidan11}
K.~M. {Pontoppidan}, G.~A. {Blake} and A.~{Smette}, \emph{\apj}, 2011,
  \textbf{733}, 84\relax
\mciteBstWouldAddEndPuncttrue
\mciteSetBstMidEndSepPunct{\mcitedefaultmidpunct}
{\mcitedefaultendpunct}{\mcitedefaultseppunct}\relax
\EndOfBibitem
\bibitem[{Hein Bertelsen} \emph{et~al.}(2014){Hein Bertelsen}, {Kamp}, {Goto},
  {van der Plas}, {Thi}, {Waters}, {van den Ancker}, and {Woitke}]{Bertelsen14}
R.~P. {Hein Bertelsen}, I.~{Kamp}, M.~{Goto}, G.~{van der Plas}, W.-F. {Thi},
  L.~B.~F.~M. {Waters}, M.~E. {van den Ancker} and P.~{Woitke}, \emph{\aap},
  2014, \textbf{561}, A102\relax
\mciteBstWouldAddEndPuncttrue
\mciteSetBstMidEndSepPunct{\mcitedefaultmidpunct}
{\mcitedefaultendpunct}{\mcitedefaultseppunct}\relax
\EndOfBibitem
\bibitem[{Dickman}(1978)]{Dickman78}
R.~L. {Dickman}, \emph{\apjs}, 1978, \textbf{37}, 407--427\relax
\mciteBstWouldAddEndPuncttrue
\mciteSetBstMidEndSepPunct{\mcitedefaultmidpunct}
{\mcitedefaultendpunct}{\mcitedefaultseppunct}\relax
\EndOfBibitem
\bibitem[{Pontoppidan} \emph{et~al.}(2008){Pontoppidan}, {Blake}, {van
  Dishoeck}, {Smette}, {Ireland}, and {Brown}]{Pontoppidan08}
K.~M. {Pontoppidan}, G.~A. {Blake}, E.~F. {van Dishoeck}, A.~{Smette}, M.~J.
  {Ireland} and J.~{Brown}, \emph{\apj}, 2008, \textbf{684}, 1323--1329\relax
\mciteBstWouldAddEndPuncttrue
\mciteSetBstMidEndSepPunct{\mcitedefaultmidpunct}
{\mcitedefaultendpunct}{\mcitedefaultseppunct}\relax
\EndOfBibitem
\bibitem[{Willacy} and {Woods}(2009)]{Willacy09}
K.~{Willacy} and P.~M. {Woods}, \emph{\apj}, 2009, \textbf{703}, 479--499\relax
\mciteBstWouldAddEndPuncttrue
\mciteSetBstMidEndSepPunct{\mcitedefaultmidpunct}
{\mcitedefaultendpunct}{\mcitedefaultseppunct}\relax
\EndOfBibitem
\bibitem[{D'Hendecourt} \emph{et~al.}(1985){D'Hendecourt}, {Allamandola}, and
  {Greenberg}]{dHendecourt85}
L.~B. {D'Hendecourt}, L.~J. {Allamandola} and J.~M. {Greenberg}, \emph{\aap},
  1985, \textbf{152}, 130--150\relax
\mciteBstWouldAddEndPuncttrue
\mciteSetBstMidEndSepPunct{\mcitedefaultmidpunct}
{\mcitedefaultendpunct}{\mcitedefaultseppunct}\relax
\EndOfBibitem
\bibitem[{Grevesse} \emph{et~al.}(2010){Grevesse}, {Asplund}, {Sauval}, and
  {Scott}]{Grevesse10}
N.~{Grevesse}, M.~{Asplund}, A.~J. {Sauval} and P.~{Scott}, \emph{\apss}, 2010,
  \textbf{328}, 179--183\relax
\mciteBstWouldAddEndPuncttrue
\mciteSetBstMidEndSepPunct{\mcitedefaultmidpunct}
{\mcitedefaultendpunct}{\mcitedefaultseppunct}\relax
\EndOfBibitem
\bibitem[{Banzatti} \emph{et~al.}(2013){Banzatti}, {Meyer}, {Pontoppidan}, and
  {Bruderer}]{Banzatti13}
A.~{Banzatti}, M.~{Meyer}, K.~{Pontoppidan} and S.~{Bruderer}, Protostars and
  Planets VI, Heidelberg, July 15-20, 2013. Poster \#2S034, 2013, p.~34\relax
\mciteBstWouldAddEndPuncttrue
\mciteSetBstMidEndSepPunct{\mcitedefaultmidpunct}
{\mcitedefaultendpunct}{\mcitedefaultseppunct}\relax
\EndOfBibitem
\bibitem[{Semenov} and {Wiebe}(2011)]{Semenov11}
D.~{Semenov} and D.~{Wiebe}, \emph{\apjs}, 2011, \textbf{196}, 25\relax
\mciteBstWouldAddEndPuncttrue
\mciteSetBstMidEndSepPunct{\mcitedefaultmidpunct}
{\mcitedefaultendpunct}{\mcitedefaultseppunct}\relax
\EndOfBibitem
\bibitem[{Pascucci} \emph{et~al.}(2013){Pascucci}, {Herczeg}, {Carr}, and
  {Bruderer}]{Pascucci13}
I.~{Pascucci}, G.~{Herczeg}, J.~S. {Carr} and S.~{Bruderer}, \emph{\apj}, 2013,
  \textbf{779}, 178\relax
\mciteBstWouldAddEndPuncttrue
\mciteSetBstMidEndSepPunct{\mcitedefaultmidpunct}
{\mcitedefaultendpunct}{\mcitedefaultseppunct}\relax
\EndOfBibitem
\bibitem[{Mandell} \emph{et~al.}(2008){Mandell}, {Mumma}, {Blake}, {Bonev},
  {Villanueva}, and {Salyk}]{Mandell08}
A.~M. {Mandell}, M.~J. {Mumma}, G.~A. {Blake}, B.~P. {Bonev}, G.~L.
  {Villanueva} and C.~{Salyk}, \emph{\apjl}, 2008, \textbf{681}, L25--L28\relax
\mciteBstWouldAddEndPuncttrue
\mciteSetBstMidEndSepPunct{\mcitedefaultmidpunct}
{\mcitedefaultendpunct}{\mcitedefaultseppunct}\relax
\EndOfBibitem
\end{mcitethebibliography}
\bibliographystyle{rsc}
}

\end{document}